\documentclass[12pt,english]{article}
\usepackage[T1]{fontenc}
\usepackage[latin1]{inputenc}
\usepackage{float}
\usepackage{amsmath}
\usepackage{graphicx}
\usepackage{amssymb}

\makeatletter

\usepackage{float}

\makeatletter

\usepackage{float}

\makeatletter
\usepackage{float}

\usepackage{slashed}

\makeatletter
\usepackage{a4}

\usepackage{epsfig}

\usepackage{cite}

\makeatother

\makeatother

\makeatother

\usepackage{babel}
\makeatother

\begin{document}

\title{Nilpotent noncommutativity and renormalization}

\author{R. Fresneda, D.~M.~Gitman and D.~V.~Vassilevich%
\thanks{On leave from V.~A.~Fock Institute of Physics, St.~Petersburg University,
Russia. E.mail:\ {\texttt{dmitry(at)dfn.if.usp.br}}%
}\\
 \textit{Instituto de Física, Universidade de São Paulo,}\\
 \textit{Caixa Postal 66318, CEP 05315-970, São Paulo, S.P., Brazil}}

\maketitle
\begin{abstract}
We analyze renormalizability properties of noncommutative (NC) theories
with a bifermionic NC parameter. We introduce a new 4-dimensional
scalar field model which is renormalizable at all orders of the loop
expansion. We show that this model has an infrared stable fixed point
(at the one-loop level). We check that the NC QED (which is one-loop
renormalizable with usual NC parameter) remains renormalizable when
the NC parameter is bifermionic, at least to the extent of one-loop
diagrams with external photon legs. Our general conclusion is that
bifermionic noncommutativity improves renormalizablility properties
of NC theories. 
\end{abstract}

\section{Introduction}

It is well known \cite{Rivasseau:2007ab} that noncommutative (NC)
field theories have renormalizability problems due to the so-called
UV/IR mixing \cite{UVIR1,UVIR2,UVIR3}. To overcome this difficulty,
one modifies the propagator by adding an oscillator term \cite{GW1,GW2,GW3}
in order to respect the Langmann-Szabo duality \cite{LS}, or by adding
a term with a negative power of the momentum \cite{Gurau:2008vd}.
Supersymmetry also improves the renormalizability properties of NC
theories (see, e.g., \cite{Girotti:2000gc}). Some versions of NC
supersymmetry (those which are based on the nonanticommutative superspace
\cite{Seiberg,Dimitrijevic:2007cu}, see also \cite{Zh,Kobayashi:2005pz})
have a nilpotent NC parameter, so that the star product terminates
at a finite order of its expansion. It was demonstrated \cite{Gitman:2007jp}
that having a nilpotent NC parameter does not necessarily imply supersymmetry.
In \cite{Gitman:2007jp} a nilpotent (bifermionic) NC parameter was
introduced in a bosonic theory, giving rise to many attractive properties
of that model. The aim of this work is to study to which extent having
a nilpotent (or bifermionic) NC parameter influences the renormalization.
We shall consider non-supersymmetric theories in order to separate
the effects of nilpotency from the effects of supersymmetry.

A suitable framework for such an analysis was suggested in \cite{Gitman:2007jp},
where it was proposed to consider a bifermionic NC parameter \begin{equation}
\Theta^{\mu\nu}=i\theta^{\mu}\theta^{\nu},\label{bifNC}\end{equation}
 where $\theta^{\mu}$ is a real constant fermion (a Grassmann odd
constant), $\theta^{\mu}\theta^{\nu}=-\theta^{\nu}\theta^{\mu}$.
Note that bifermionic constants appear naturally in pseudoclassical
models of relativistic particles \cite{Gitman:1994rt,Fresneda:2003gy}.
Due to the anticommutativity of $\theta^{\mu}$, the expansion of
the usual Moyal product terminates at the second term, \begin{equation}
f_{1}\star f_{2}=\exp\left(\frac{i}{2}\Theta^{\mu\nu}\partial_{\mu}^{x}\partial_{\nu}^{y}\right)f_{1}(x)f_{2}(y)\vert_{y=x}=f_{1}\cdot f_{2}-\frac{1}{2}\theta^{\mu}\theta^{\nu}\partial_{\mu}f_{1}\partial_{\nu}f_{2}.\label{bifstar}\end{equation}
 The star-product, therefore, becomes local.

In \cite{Gitman:2007jp} a bifermionic NC parameter was used to construct
a two-dimensional field theory model which, in contrast to usual time-space
NC models, has a locally conserved energy momentum tensor, a well-defined
conserved Hamiltonian, and can be canonically quantized without any
difficulties. Besides, the model appears to be renormalizable. In
the present work we study whether bifermionic noncommutativity helps
renormalize theories in four dimensions.

First we explore a model which is a four-dimensional version of the
model suggested in \cite{Gitman:2007jp} (this is nothing else than
NC $\varphi^{4}$ with an additional interaction included to make
it less trivial). We find that for a bifermionic NC parameter this
model becomes renormalizable at all orders of the loop expansion.
We also study the one-loop renormalization group equations and find
an infrared stable fixed point where all couplings vanish.

From the technical point of view, having a bifermionic NC parameter
looks similar to expanding the theory in $\Theta$ and keeping just
a few leading terms. The ultraviolet properties of the expanded and
full theories are rather different, and, sometimes, expanded theories
behave worse (see, e.g., \cite{Wulkenhaar:2001sq}). The reason is
that, on one hand, the propagator in expanded theories does not have
an oscillatory behavior, and, on the other hand, dangerous momentum-dependent
vertices appear. All these problems appear also in theories with bifermionic
noncommutativity, but there is also an effect which improves the ultraviolet
behavior. Namely, some divergent terms vanish due to $\theta^{2}=0$.
Here we take the NC QED (which is one-loop renormalizable if the standard
NC parameter is used) and demonstrate that with a bifermionic NC parameter
this model remains renormalizable at least for one loop diagrams with
external photons.

\section{A scalar field model}

The action of the model we consider in this section reads \begin{eqnarray}
 &  & S=\int d^{4}x\left(\frac{1}{2}(\partial_{\mu}\varphi_{1})^{2}+\frac{1}{2}(\partial_{\mu}\varphi_{2})^{2}+\frac{1}{2}(\partial_{\mu}\varphi)^{2}-\frac{1}{2}m_{1}^{2}\varphi_{1}^{2}-\frac{1}{2}m_{2}^{2}\varphi_{2}^{2}-\frac{1}{2}m^{2}\varphi^{2}\right.\nonumber \\
 &  & \qquad\qquad\left.-\frac{ei}{2}[\varphi_{1},\varphi_{2}]_{\star}\star\varphi\star\varphi-\frac{\lambda}{24}\varphi_{\star}^{4}\right),\label{act1}\end{eqnarray}
 which is a four dimensional version of a model suggested in \cite{Gitman:2007jp}.
The motivations for taking this particular form of the model are as
follows. Since any symmetrized star product with a bifermionic parameter
is equivalent to the usual commutative pointwise product, we need
at least two fields, $\varphi_{1}$ and $\varphi_{2}$, to construct
a non-trivial polynomial interaction term. As was explained in \cite{Gitman:2007jp},
even two fields are not enough, so we take another scalar field $\varphi$
to construct the interaction term with a coupling constant $e$. We
also added a self-interaction term $\varphi_{\star}^{4}=\varphi\star\varphi\star\varphi\star\varphi$
to make the dynamics more interesting. $e$ and $\lambda$ are real
coupling constants.

In \cite{Gitman:2007jp} it was demonstrated that a two dimensional
model with the same Lagrange density as in (\ref{act1}) is renormalizable.
It is relatively easy to achieve renormalizability in two dimensions.
For example, there is a model of NC gravity in two dimensions for
which the entire quantum generating functional of Green functions
can be calculated non-perturbatively at all orders of the loop expansion
\cite{Vassilevich:2004ym} by using methods developed earlier in the
commutative case \cite{Kummer:1996hy}. Here, to be closer to physics,
we consider a four-dimensional model (\ref{act1}).

Due to our choice (\ref{bifNC}) of the NC parameter, the interaction
part of the action (\ref{act1}) looks rather simple, \begin{equation}
S_{{\rm int}}=\int d^{4}x\left(\frac{ei}{2}(\theta^{\mu}\partial_{\mu}\varphi_{1})(\theta^{\nu}\partial_{\nu}\varphi_{2})\varphi^{2}-\frac{\lambda}{24}\varphi^{4}\right).\label{Sint}\end{equation}

Now we are ready to derive the Feynman rules for our model. The propagators
are the standard propagators of massive scalar fields. There are two
vertices, the standard $\varphi^{4}$ vertex and a new vertex, which
depends on the NC parameter.

\begin{figure}[H]
 $\qquad$\includegraphics[scale=0.4]{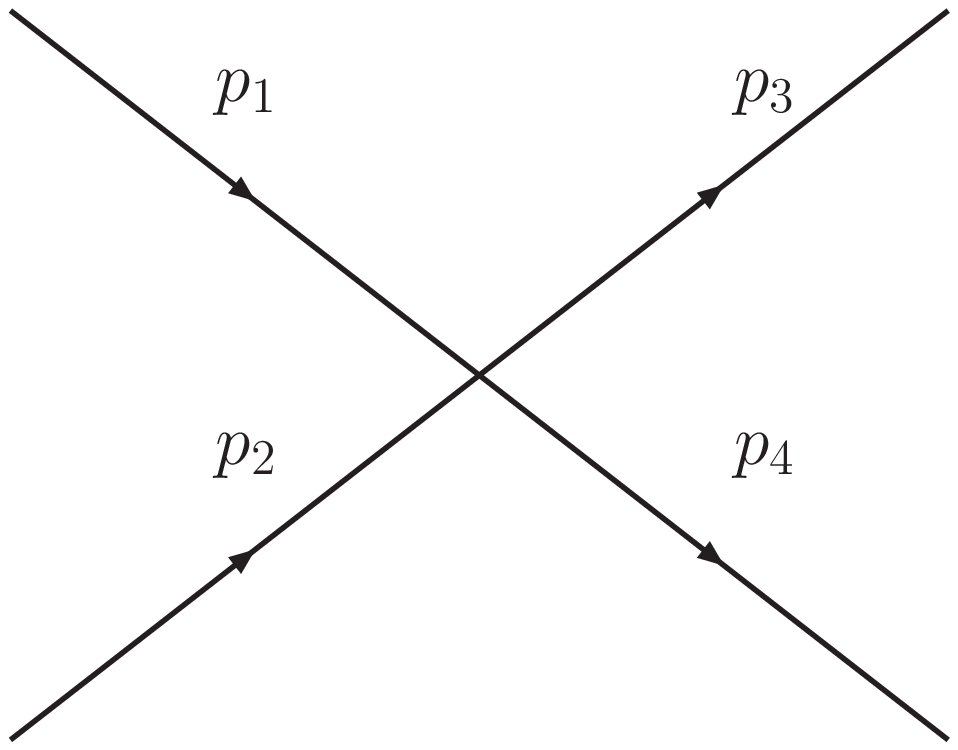}$\qquad$\includegraphics[scale=0.4]{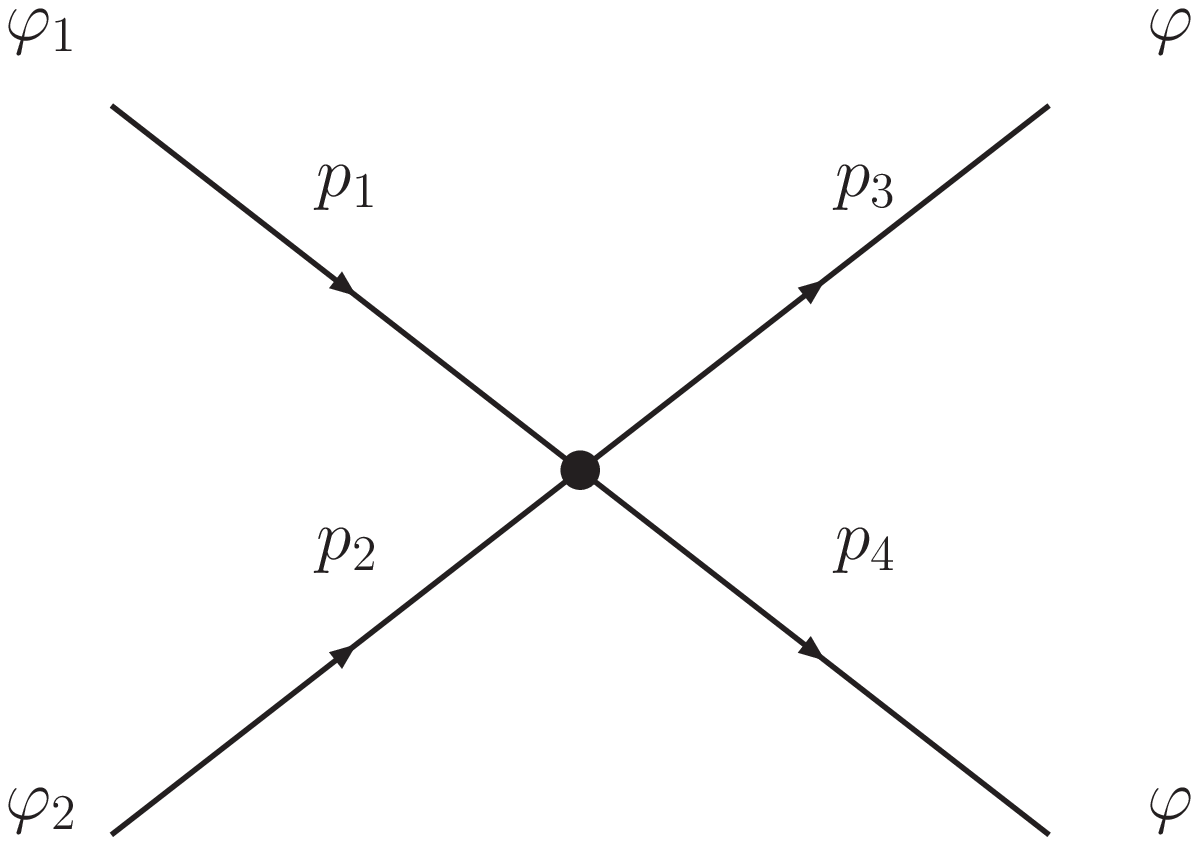}

\caption{The standard $\varphi^{4}$ vertex and the new vertex $-\frac{ie}{2}\theta p_{1}\theta p_{2}$.}

\label{fig:phi4} 
\end{figure}

\begin{figure}[H]
 $\qquad$\includegraphics[scale=0.4]{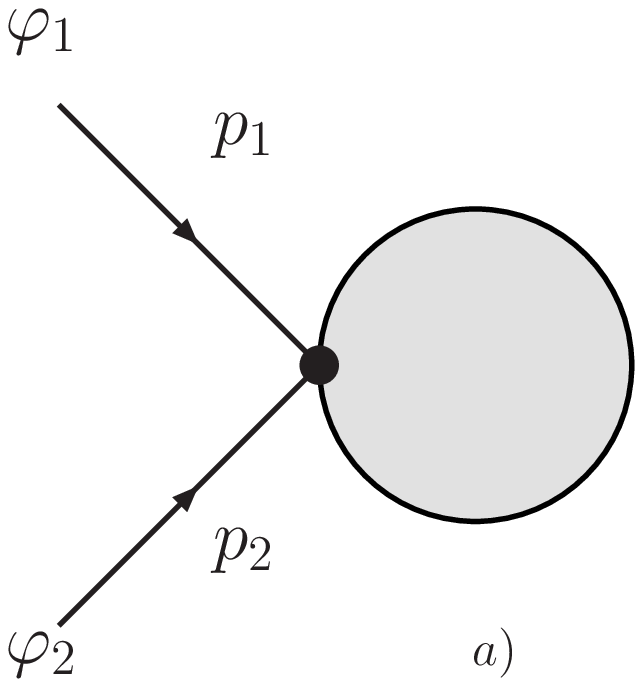}$\qquad$\includegraphics[scale=0.4]{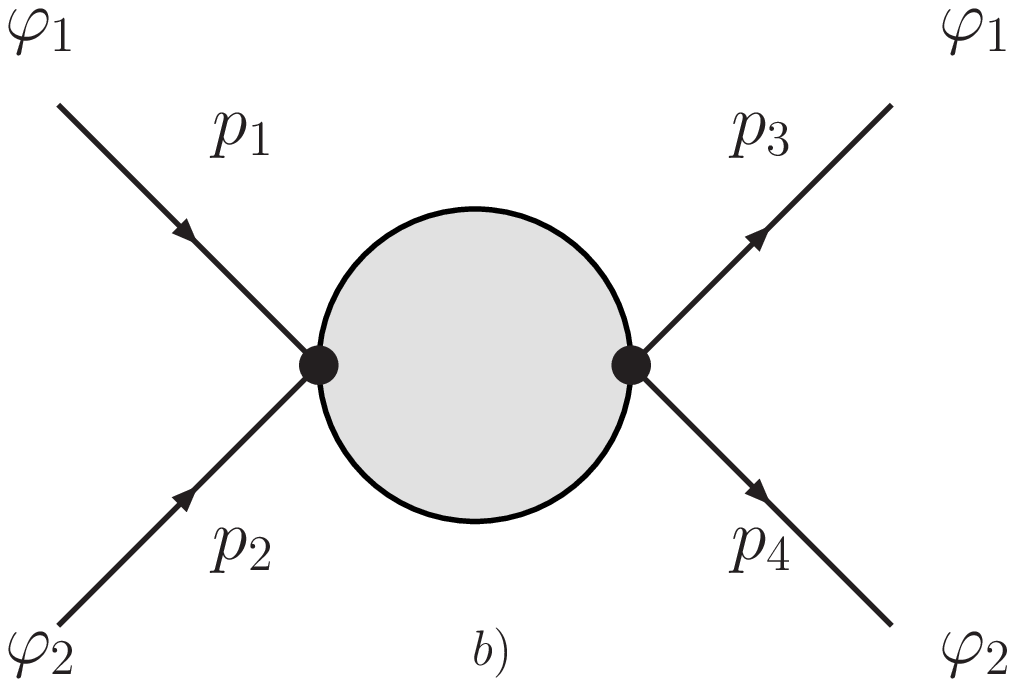}$\qquad$\includegraphics[scale=0.4]{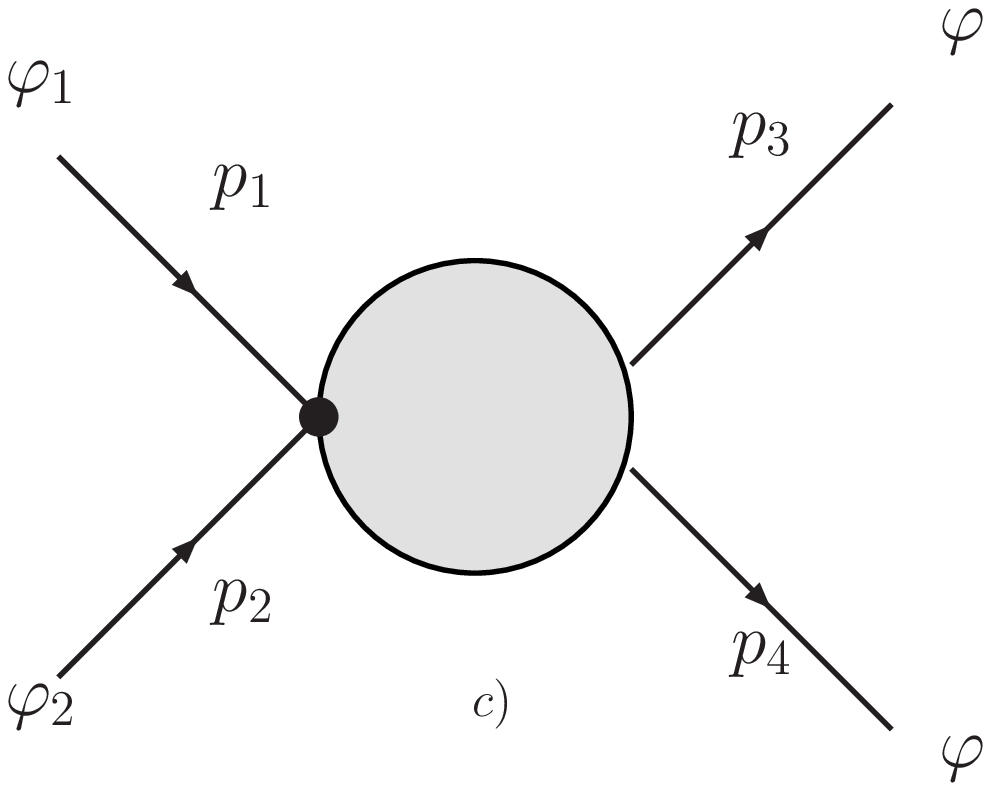}

\caption{The three divergent diagrams.}

\label{fig:divdiag} 
\end{figure}

The main observation which proves the renormalizability of (\ref{act1})
is that any diagram with an \textit{internal} line of either $\varphi_{1}$
or $\varphi_{2}$ field vanishes. Indeed, any internal line of these
fields inevitably connects two \char`\"{}new\char`\"{} vertices and,
therefore, receives a multiplier $(\theta\cdot k)^{2}=0$, where $k$
is the corresponding momentum. Power-counting renormalizability of
our model follows then by standard arguments, precisely as in the
commutative case. Consider a diagram with $N$ vertices and $2K$
external legs. This diagram has $\frac{1}{2}(4N-2K)=2N-K$ internal
lines, giving the total power of the momenta in the integrand $-2(2N-K)$.
The momenta of the internal lines are restricted by $N-1$ delta-functions,
where $-1$ corresponds to conservation of the total momenta of all
external legs. Putting all this together, we obtain that the degree
of divergence is $4-2K$, as in the commutative $\varphi^{4}$ theory.
The power-counting divergent diagrams are the ones with $2$ or $4$
external legs. The diagrams containing $\varphi$ legs only are precisely
the same as in the commutative case, and they are renormalized in
precisely the same way. Let us consider the diagrams with $\varphi_{1}$
and $\varphi_{2}$ legs. There are three types of such diagrams (see
Fig. \ref{fig:divdiag})

The diagram on Fig. \ref{fig:divdiag}a is proportional to $(p\theta)^{2}$,
and, therefore, vanishes. The diagram on Fig. \ref{fig:divdiag}b
contains $(p_{1}\theta)(p_{2}\theta)(p_{3}\theta)(p_{4}\theta)=0$,
due to momentum conservation, $p_{1}+p_{2}=p_{3}+p_{4}$. The diagram
of Fig. \ref{fig:divdiag}c is at most logarithmically divergent.
Therefore, their divergent parts are proportional to the lowest power
of the external momenta, i.e., to $(p_{1}\theta)(p_{2}\theta)$. It
is easy to see, that such divergences can be removed by a renormalization
of the coupling $e$ in the action (\ref{act1}). We conclude that
the model (\ref{act1}) with a bifermionic NC parameter is renormalizable
at all orders of the loop expansion.

The renormalization of all parameters related to the field $\varphi$
(the renormalization of $m$, $\lambda$ and of the wave function
$\varphi$) is not sensitive to the presence of the other fields $\varphi_{1}$
and $\varphi_{2}$. There is no renormalization of the mass or of
the wave function $\varphi_{1}$ or $\varphi_{2}$. By comparing combinatoric
factors appearing in front of the relevant Feynman diagrams, and using
the standard result \cite{Peskin:1995ev} for commutative $\varphi^{4}$
theory in the dimensional regularization scheme, one can derive a
relation \begin{equation}
3\frac{\delta e}{e}=\frac{\delta\lambda}{\lambda}=\frac{\lambda}{16\pi^{2}}\frac{3}{\epsilon}\label{dedl}\end{equation}
 between infinite one-loop renormalizations of the charges $e$ and
$\lambda$. The $\beta$-function for $\lambda$ is well known \cite{Peskin:1995ev}
\begin{equation}
\beta_{\lambda}=-\epsilon\lambda+\frac{3\lambda^{2}}{16\pi^{2}}+O(\lambda^{3}).\label{betalam}\end{equation}
 From the relation (\ref{dedl}) one can obtain the anomalous dimension
of the coupling $e$, $\beta_{e}$, using the fact that the bare coupling
is renormalization-group invariant,\[
\mu\frac{de_{0}}{d\mu}=0\,,\,\, e_{0}=\mu^{\epsilon}e\left(1+\frac{\lambda}{16\pi^{2}}\frac{1}{\epsilon}\right)\,.\]
 Explicitly,\begin{eqnarray*}
\mu\frac{d}{d\mu}e_{0} & = & \mu^{\epsilon}\left(\epsilon e+\frac{e\lambda}{16\pi^{2}}\right)+\mu^{\epsilon}\left[\beta_{e}\left(1+\frac{\lambda}{16\pi^{2}}\frac{1}{\epsilon}\right)+\frac{e}{16\pi^{2}}\frac{1}{\epsilon}\beta_{\lambda}\right]=0\,,\end{eqnarray*}
 which implies\begin{eqnarray*}
\beta_{e} & = & -\left[\epsilon e+\frac{e\lambda}{16\pi^{2}}+\frac{e}{16\pi^{2}}\frac{1}{\epsilon}\beta_{\lambda}\right]\left(1-\frac{\lambda}{16\pi^{2}}\frac{1}{\epsilon}\right)\\
 & = & -\epsilon e+\frac{\lambda e}{16\pi^{2}}+O\left(e\lambda^{2}\right).\end{eqnarray*}

Now we can remove the regularization by setting $\epsilon=0$ and
solve the renormalization group equations \begin{equation}
\mu\frac{d}{d\mu}\lambda(\mu)=\beta_{\lambda}(\lambda(\mu)),\qquad\mu\frac{d}{d\mu}e(\mu)=\beta_{e}(e(\mu))\label{RGeq}\end{equation}
 for the running couplings $\lambda(\mu)$ and $e(\mu)$. The initial
conditions are $\lambda(\mu_{0})=\lambda$, $e(\mu_{0})=e$ with $\mu_{0}$
being a normalization scale. Since $\beta_{\lambda}$ does not depend
on $e$, the equation for $\lambda(\mu)$ may be solved first, giving
the well-known result \begin{equation}
\lambda\left(\mu\right)=\lambda\left(1-\frac{3}{16\pi^{2}}\lambda\ln\frac{\mu}{\mu_{0}}\right)^{-1}.\label{lamu}\end{equation}
 Solving then the equation for $e(\mu)$ we obtain \begin{equation}
e\left(\mu\right)=e\left(1-\frac{3\lambda}{16\pi^{2}}\ln\frac{\mu}{\mu_{0}}\right)^{-\frac{1}{3}}.\label{emu}\end{equation}
 In the limit $\mu\to0$ both couplings vanish, and we have an infrared
stable fixed point. Note, that $e(\mu)$ vanishes slower than $\lambda(\mu)$
while approaching the fixed point.

\section{Noncommutative QED with bifermionic parameter}

Let us consider NC QED in Euclidean space with the classical action
\begin{equation}
S_{cl}=\int d^{4}x\left[\frac{1}{4g^{2}}\hat{F}_{\mu\nu}^{2}+\bar{\psi}\left(i\gamma_{\mu}D_{\mu}\right)\psi\right]\label{clac}\end{equation}
 where $D_{\mu}\psi=\partial_{\mu}\psi-iA_{\mu}\star\psi$ and \[
\hat{F}_{\mu\nu}=F_{\mu\nu}-i(A_{\mu}\star A_{\nu}-A_{\nu}\star A_{\mu}),\qquad F_{\mu\nu}=\partial_{\mu}A_{\nu}-\partial_{\nu}A_{\mu}\,.\]
 The $\gamma$-matrices satisfy $\left\{ \gamma_{\mu},\gamma_{\nu}\right\} =2\delta_{\mu\nu}$
and are hermitian, $\delta_{\mu\nu}=\mathrm{diag}\left(1,1,1,1\right)$.
For ordinary NC parameter, this theory is known to be one-loop renormalizable
\cite{Hayakawa:1999yt,Hayakawa:1999zf}. But an expansion in $\Theta$
can violate renormalizability already at one loop, as was demonstrated
in \cite{Wulkenhaar:2001sq} in the framework of the Seiberg-Witten
map.

Here we check whether NC QED remains renormalizable at one loop if
the NC parameter is bifermionic (\ref{bifNC}). To simplify our analysis
we consider the case when only $\psi$ is quantized while $A_{\mu}$
remains a classical background field. One can check that this corresponds
to retaining all diagrams with external photons in the Lorentz gauge.
Renormalizability in such a simplified model means that the one-loop
divergence is proportional to the corresponding term in the classical
action (\ref{clac}), namely, to $\hat{F}_{\mu\nu}^{2}$. The effective
action can be formally written as \begin{equation}
W=-\ln\det\slashed{D}=-\frac{1}{2}\ln\det\slashed{D}^{2}\label{effact}\end{equation}
 where $\slashed{D}$ is the Dirac operator on noncommutative $\mathbb{R}^{4}$
in the presence of an external electromagnetic field. \begin{equation}
\slashed{D}=i\gamma_{\mu}\left(\partial_{\mu}-iA_{\mu}\star\right)=i\gamma_{\mu}\left(\partial_{\mu}-iA_{\mu}+\frac{i}{2}\theta\partial A_{\mu}\theta\partial\right)\,,\,\,\theta\partial\equiv\theta_{\mu}\partial_{\mu}\,.\label{Dirac}\end{equation}
 To avoid writing too many brackets we adopt the convention that the
derivative only acts on the function which is next to it on the right
(ignoring, of course, any number of $\theta$'s or other derivatives
which may appear in between). For example, $\theta\partial A_{\mu}\theta\partial=(\theta\partial A_{\mu})\theta\partial$
is a first-order differential operator.

It is convenient to use the zeta-function regularization of functional
determinants \cite{Dowker:1975tf,Hawking:1976ja}, so that the regularized
effective action (\ref{effact}) reads $W^{{\rm reg}}=\frac{1}{2}\zeta(\slashed{D}^{2},s)\Gamma(s)$
where $\zeta(\slashed{D}^{2},s)={\rm Tr}_{L^{2}}((\slashed{D}^{2})^{-s})$.
In the physical limit, $s\to0$, the regularized effective action
diverges, and the divergent part reads \begin{equation}
W^{{\rm div}}=\frac{1}{2s}\zeta(\slashed{D}^{2},0).\label{Wdiv}\end{equation}
 Usually, $\slashed{D}^{2}$ is an operator of Laplace type, so that
the heat trace \begin{equation}
K(\slashed{D}^{2};t)={\rm Tr}_{L^{2}}(e^{-t\slashed{D}^{2}})\label{heatr}\end{equation}
 exists and admits an asymptotic expansion \begin{equation}
K(\slashed{D}^{2};t)\simeq\sum_{k\geq0}t^{\left(k-n\right)/2}a_{k}\left(\slashed{D}^{2}\right)\label{asymptotex}\end{equation}
 as $t\to+0$. Here $n$ the is dimension of the underlying manifold.
A review of the heat kernel expansion can be found in \cite{Vassilevich:2003xt}
for commutative manifolds, and in \cite{Vassilevich:2007fq} for the
NC case. Let us assume that the expansion (\ref{asymptotex}) is valid
for the operator (\ref{Dirac}). (This will be demonstrated in a moment).
Then, by using the Mellin transform, one can show \begin{equation}
\zeta(\slashed{D}^{2},0)=a_{4}(\slashed{D}^{2})\label{zea4}\end{equation}
 in $n=4$ dimensions. There is no good spectral theory for differential
operators with symbols depending on fermionic parameters. To be on
the safe side, we shall evaluate (\ref{zea4}) by two independent
methods.

First, we use existing results on the heat kernel expansion on NC
manifolds. The operator \begin{equation}
\slashed{D}^{2}=-\left((\partial_{\mu}-iA_{\mu}\star)^{2}-\frac{i}{4}[\gamma^{\mu},\gamma^{\nu}]\hat{F}_{\mu\nu}\star\right),\label{sDstar}\end{equation}
 (where partial derivatives act all the way to the right), has left
star-multiplications only (meaning that in the eigenvalue equation
$\slashed{D}^{2}\psi=\lambda\psi$ all background fields multiply
$\psi$ from the left), and, therefore, falls into the category considered
in \cite{Vassilevich:2003yz,Gayral:2004ww}. The calculations made
in \cite{Vassilevich:2003yz}%
\footnote{The paper \cite{Vassilevich:2003yz} treated the case of a NC torus,
and the case of a NC plane was done in \cite{Gayral:2004ww}. In the
present context distinctions between the torus and the plane are not
essential.%
} are regular at $\Theta=0$ and survive an expansion to a finite order
in $\Theta$ (see eqs.(15) - (26) there). Note that such a statement
is not true for operators having both right and left star multiplications
\cite{Vassilevich:2005vk,Gayral:2006vd}. Anyway, we are allowed to
use the results of \cite{Vassilevich:2003yz,Gayral:2004ww} for the
operator (\ref{sDstar}). First, we bring $\slashed{D}^{2}$ to the
standard form \begin{equation}
\slashed{D}^{2}=-\left(\hat{\nabla}_{\mu}\hat{\nabla}_{\mu}+\hat{E}\star\right),\qquad\hat{\nabla}_{\mu}\equiv\partial_{\mu}+\hat{\omega}_{\mu}\star\,,\label{Dhat}\end{equation}
 where \begin{equation}
\hat{\omega}_{\mu}=-iA_{\mu},\qquad\hat{E}=-\frac{i}{4}[\gamma^{\mu},\gamma^{\nu}]\hat{F}_{\mu\nu}.\label{oEhat}\end{equation}
 Then, according to \cite{Vassilevich:2003yz,Gayral:2004ww}, the
asymptotic expansion (\ref{asymptotex}) exists and the coefficient
$a_{4}$ reads \begin{equation}
a_{4}=\frac{1}{(4\pi)^{2}}\,\frac{1}{12}\int d^{4}x\,{\rm tr}\,(6\hat{E}\star\hat{E}+\hat{\Omega}_{\mu\nu}\star\hat{\Omega}_{\mu\nu})\label{a4hat}\end{equation}
 with $\hat{\Omega}_{\mu\nu}=[\hat{\nabla}_{\mu},\hat{\nabla}_{\nu}]$.
By substituting (\ref{oEhat}) in (\ref{a4hat}) and taking the trace,
we obtain \begin{equation}
a_{4}(\slashed{D}^{2})=\frac{1}{\left(4\pi\right)^{2}}\frac{2}{3}\int d^{4}x\hat{F}_{\mu\nu}\star\hat{F}_{\mu\nu}\,.\label{a4Dstar}\end{equation}

The other method we use does not rely on the star-product structure,
but rather uses an expanded form of the operator \begin{eqnarray}
 &  & \slashed{D}^{2}=-(\partial^{2}-2iA_{\mu}\partial_{\mu}-i(\partial_{\mu}A_{\mu})-A^{2})-i\left(\theta\partial\right)A^{\mu}\left(\theta\partial\right)\partial_{\mu}\nonumber \\
 &  & \qquad-\frac{i}{8}\left[\gamma^{\mu},\gamma^{\nu}\right]\left(\theta\partial\right)F_{\mu\nu}\left(\theta\partial\right)-\frac{i}{2}\left(\theta\partial\right)\partial^{\mu}A_{\mu}\left(\theta\partial\right)-A^{\mu}\left(\theta\partial\right)A_{\mu}\left(\theta\partial\right)\nonumber \\
 &  & \qquad-\frac{1}{4}\left[\gamma^{\mu},\gamma^{\nu}\right]\left(\theta\partial\right)A_{\mu}\left(\theta\partial\right)A_{\nu}+\frac{i}{4}\left[\gamma^{\mu},\gamma^{\nu}\right]F_{\mu\nu}\,.\label{D2exp}\end{eqnarray}
 The coefficient $a_{4}$ can be read off from the seminal paper by
Gilkey \cite{Gilkey:1975iq} by identifying corresponding invariants.
For any Laplace type operator of the form \begin{equation}
P=-\left(g^{\mu\nu}\partial_{\mu}\partial_{\nu}+a^{\sigma}\partial_{\sigma}+b\right)\label{Pop}\end{equation}
 one identifies $g^{\mu\nu}$ with a Riemannian metric (to enable
such an identification the leading symbol must be a unit matrix in
spinorial indices - a property which is fortunately true for the operator
(\ref{D2exp})). There is a unique connection $\omega$ such that
$P$ may be presented as \begin{equation}
P=-\left(g^{\mu\nu}\nabla_{\mu}\nabla_{\nu}+E\right),\label{Pstan}\end{equation}
 where the covariant derivative $\nabla=\nabla^{[R]}+\omega$ contains
the Riemann connection and a gauge part. The zeroth-order part reads
$E=b-g^{\mu\nu}\left(\partial_{\mu}\omega_{\nu}+\omega_{\nu}\omega_{\mu}-\omega_{\sigma}\Gamma_{\nu\mu}^{\sigma}\right)$,
where $\Gamma_{\nu\mu}^{\sigma}$ is the Christoffel symbol of the
metric $g^{\mu\nu}$. One also introduces the field strength tensor
$\Omega_{\mu\nu}=\partial_{\mu}\omega_{\nu}-\partial_{\nu}\omega_{\mu}+\left[\omega_{\mu},\omega_{\nu}\right]$.

In $n=4$ the relevant heat kernel coefficient reads \begin{equation}
a_{4}\left(P\right)=\frac{1}{(4\pi)^{2}}\,\frac{1}{12}\int d^{4}x\sqrt{g\left(x\right)}\mathrm{tr}\left(6E^{2}+\Omega_{\mu\nu}\Omega_{\rho\sigma}g^{\mu\rho}g^{\nu\sigma}+[R^{2}-\mathrm{terms}]\right)\,.\end{equation}
 The terms quadratic in the Riemann curvature tensor are not written
explicitly. The model was initially formulated in flat Euclidean space,
so that there are no distinctions between upper and lower indices.
Whenever we need to contract a pair of indices with the effective
metric $g^{\mu\nu}$, the metric is written explicitly.

Let us restrict ourselves to the terms which are of zeroth and second
order in $\theta$. From eq.(\ref{D2exp}) one can read off the metric
$g^{\mu\nu}$ \begin{equation}
g^{\mu\nu}=\delta^{\mu\nu}+\frac{i}{2}\theta\partial\left(A^{\mu}\theta^{\nu}+A^{\nu}\theta^{\mu}\right)\,,\,\, g_{\mu\nu}=\delta_{\mu\nu}-\frac{i}{2}\theta\partial\left(A_{\mu}\theta_{\nu}+A_{\nu}\theta_{\mu}\right),\label{gmunu}\end{equation}
 the Christoffel symbol \[
\Gamma_{\nu\sigma}^{\mu}=\frac{i}{4}\delta^{\mu\kappa}\theta\partial\left[\theta_{\sigma}F_{\kappa\nu}+\theta_{\nu}F_{\kappa\sigma}-\theta_{\kappa}(\partial_{\sigma}A_{\nu}+\partial_{\nu}A_{\sigma})\right]\,,\]
 and $a^{\mu}$ and $b$, \begin{align*}
 & a^{\mu}=\frac{i}{8}\left[\gamma_{\kappa},\gamma_{\nu}\right]\theta\partial F_{\kappa\nu}\theta^{\mu}+\frac{i}{2}\theta\partial\partial_{\nu}A_{\nu}\theta^{\mu}+A_{\nu}\theta\partial A_{\nu}\theta^{\mu}-2iA^{\mu}\\
 & b=\frac{1}{4}\left[\gamma_{\mu},\gamma_{\nu}\right]\theta\partial A_{\mu}\theta\partial A_{\nu}-\frac{i}{4}\left[\gamma_{\mu},\gamma_{\nu}\right]F_{\mu\nu}-i\partial A-A^{2}\,.\end{align*}
 From these expressions we calculate the gauge connection\begin{align*}
\omega_{\mu} & =\frac{1}{2}g_{\mu\nu}\left(a^{\nu}+g^{\kappa\sigma}\Gamma_{\kappa\sigma}^{\nu}\right)\\
 & =-iA_{\mu}-\frac{1}{2}\left(\theta\partial\right)A_{\mu}\left(\theta A\right)+\frac{i}{4}\left(\theta\partial\right)\theta^{\kappa}F_{\mu\kappa}+\frac{i}{16}\left[\gamma^{\kappa},\gamma^{\sigma}\right]\theta\partial F_{\kappa\sigma}\theta_{\mu}\,,\end{align*}
 and the trace of $E^{2}$ and $\Omega^{2}$ follow \begin{align*}
 & \mathrm{tr}E^{2}=2\hat{F}_{\mu\nu}\hat{F}_{\mu\nu}+2iF_{\mu\nu}\left(\theta A\right)\left(\theta\partial\right)F_{\mu\nu}\,,\,\,\hat{F}_{\mu\nu}=F_{\mu\nu}+i\left(\theta\partial\right)A_{\mu}\left(\theta\partial\right)A_{\nu}\,,\\
 & \mathrm{tr}g^{\mu\kappa}g^{\nu\sigma}\Omega_{\mu\nu}\Omega_{\kappa\sigma}=-4\hat{F}_{\mu\nu}\hat{F}_{\mu\nu}+4iF_{\mu\nu}\left(\theta\partial\right)F_{\mu\nu}\left(\theta A\right)\,.\end{align*}

The Riemann tensor for the metric (\ref{gmunu}) is at least of the
second order in $\theta$. Therefore, the curvature square terms are
at least of the fourth order in $\theta$ and must be neglected.

Finally, we are able to compute $a_{4}$, \[
a_{4}\left(\slashed{D}^{2}\right)=\frac{1}{\left(4\pi\right)^{2}}\frac{2}{3}\int d^{4}x\hat{F}_{\mu\nu}\hat{F}_{\mu\nu}\,,\]
 which is in agreement with (\ref{a4Dstar}).

The two methods we used above to calculate the heat kernel coefficient
$a_{4}$ differ in the way we treated derivatives contained in the
star product. In the second method these derivatives modify the first
and the second order parts of the corresponding differential operator,
and, therefore, the effective metric and the effective connections
are changed. According to the first method, the star-product as a
whole is considered as a multiplication, i.e., as a zeroth order operator.
This ensures regularity of the heat kernel expansion \cite{Vassilevich:2003yz,Gayral:2004ww}
for small $\Theta$. For more general NC Laplacians (containing both
right and left star multiplications) this regularity is lost \cite{Vassilevich:2005vk,Gayral:2006vd}.
However, let us consider the heat operator $h(t)=e^{-t(P_{0}+P_{2})}$
where $P_{0}$ does not depend on $\theta$, while $P_{2}$ is at
least bilinear in the (fermionic) parameter. Obviously, $h(t)$ can
be expanded in series in $P_{2}$, and convergence is not an issue,
since the expansion terminates. These simple arguments show that in
a more general case the second method will probably work, while the
first one will probably not.

By collecting together (\ref{Wdiv}), (\ref{zea4}) and (\ref{a4Dstar}),
we see that the divergent part of the effective action is proportional
to $\hat{F}_{\mu\nu}^{2}$ and may be cancelled by a renormalization
of the coupling $g$ in the classical action (\ref{clac}). Therefore,
the model (\ref{clac}) with quantized spinor and background vector
fields is renormalizable.

\section{Conclusions}

In this paper we have studied the renormalization properties of NC
theories in four dimensions with a bifermionic NC parameter. We have
found a scalar model which is renormalizable at all orders of the
loop expansion, thus adding a new example to a (not very rich) family
of renormalizable non-supersymmetric NC theories in four dimensions.
We have also found that this model has an infrared stable fixed point
at the one-loop level.

We also took another model, the NC QED, which is one-loop renormalizable
with the usual NC parameter, and checked that the introduction of
a bifermionic NC parameter does not destroy the one-loop renormalizability
at least in the sector with external photon legs. We conclude that
bifermionic noncommutativity is renormalization-friendly. Thus it
seems to be a rather promising version of noncommutativity, worth
being taken seriously, and prompting further studies.

\section*{Acknowledgments}

The work of D.M.G.\ and D.V.V.\ was supported in part by FAPESP
and CNPq. R.F wishes to thank FAPESP for support. The work of R.F
was fully supported by FAPESP.

\end{document}